\documentclass[a4paper, 10pt, twocolumn]{article}

\usepackage{amsfonts,amssymb,amsmath,amsthm}
\usepackage{subfigure}
\usepackage{graphicx}
\usepackage[footnotesize]{caption}

\usepackage[top=1.5cm, left=1.5cm, right=1.5cm, bottom=1.5cm]{geometry}

\makeatletter
\let\origsubsection\subsection
\renewcommand\subsection{\@ifstar{\starsubsection}{\nostarsubsection}}

\newcommand\nostarsubsection[1]
{\subsectionprelude\origsubsection{#1}\subsectionpostlude}

\newcommand\starsubsection[1]
{\subsectionprelude\origsubsection*{#1}\subsectionpostlude}

\newcommand\subsectionprelude{%
  \vspace{-2mm}
}

\newcommand\subsectionpostlude{%
  \vspace{-1mm}
}
\makeatother

\newcommand{\vse}{\vspace{-1.5mm}}

\usepackage{hyperref,xspace}

\newtheorem{theorem}{Theorem}[]

\DeclareMathOperator{\signc}{sign_{\bb C}}

\DeclareMathOperator{\diag}{diag}

\DeclareMathOperator*{\argmin}{arg\min}

\newcommand{\bs}{\boldsymbol}
\newcommand{\bb}{\mathbb}
\newcommand{\cl}{\mathcal}

\newcommand{\nx}{x^\star}
\newcommand{\scp}[2]{\langle #1,\, #2 \rangle}

\newcommand{\ts}{\textstyle}

\newcommand{\ie}{\emph{i.e.},\xspace}

\newcommand{\im}{{\sf i}}

\newcommand{\iid}{%
    \ifmmode% math mode
        \mathrm{i.i.d.}%
    \else%
        i.i.d.\@\xspace%
    \fi%
}

\newcommand{\st}{%
    \ifmmode% math mode
        \quad\mathrm{s.t.}\quad%
    \else%
        s.t.\@\xspace%
    \fi%
}

\newcommand{\whp}{\mbox{w.h.p.\@}\xspace}

\title{\vspace{-5mm}\emph{Keep the phase}! Signal recovery in phase-only compressive sensing\vspace{-2mm}}

\author{L. Jacques$^{1,*}$ and T. Feuillen$^1$.\\
\footnotesize $^1$UCLouvain, Louvain-la-Neuve, Belgium.\ $^*$Corresponding author. 
\vspace{-5mm}}
\date{\empty} % no need for a date

\renewenvironment{abstract}{\bf\small {\em\ Abstract---}}{}

\begin{document}

\maketitle

\begin{abstract}
We demonstrate that a sparse signal can be estimated from the phase of complex random measurements, in a \emph{phase-only compressive sensing} (PO-CS) scenario. With high probability and up to a global unknown amplitude, we can \emph{perfectly} recover such a signal if the sensing matrix is a complex Gaussian random matrix and the number of measurements is large compared to the signal sparsity. Our approach consists in recasting the (non-linear) PO-CS scheme as a linear compressive sensing model. We built it from a signal normalization constraint and a phase-consistency constraint. Practically, we achieve stable and robust signal direction estimation from the basis pursuit denoising program. Numerically, robust signal direction estimation is reached at about twice the number of measurements needed for signal recovery in compressive sensing.\footnote{This work is a 2-page summary restricting \cite{LJ20} to the case of sparse signals.}   
\end{abstract}

\subsection{Introduction}
\label{sec:introduction}

Forty years ago, Oppenheim and collaborators~\cite{Oppenheim_1982,Oppenheim_1981} determined that, under certain conditions and up to a global unknown amplitude, one can practically reconstruct a band-limited signal from the phase of its Fourier transform.

Later, Boufounos~\cite{boufounos2013sparse} considered a similar question in the context of complex compressive sensing where measurements of a sparse signal are obtained from its multiplication with a complex, fat sensing matrix. The author proved that, in the case of complex Gaussian random sensing matrices, one can estimate the direction of such a signal from the phase of the compressive measurements (or \emph{phase-only} compressive sensing -- PO-CS). PO-CS naturally extends one-bit compressive sensing, which, in the real field, keeps only the sign of real random projections~\cite{boufounos2008,jacques2013robust,plan2012robust}. The estimation error then provably decays when the number of measurements increases. Boufounos also designed in~\cite{boufounos2008} a greedy algorithm whose estimate is both sparse and phase-consistent with the unknown signal --- this signal and the estimate share identical phases in the random projection domain. Surprisingly, this specific algorithm succeeds numerically in \emph{perfectly} estimating the observed signal direction when the number of measurements is large enough; a fact that is not explained by the theory. 

With this work, we first demonstrate that, in a noiseless scenario, perfect estimation of a sparse signal direction from the phase of its complex Gaussian random measurements is possible. This result strikingly differs from known reconstruction guarantees in the context of (real) one-bit CS~\cite{boufounos2008,jacques2013robust,plan2012robust}; in this case, the direction of a low-complexity signal can only be estimated up to a lower-bounded error~\cite{jacques2013robust}.  

Second, we show that one can practically reconstruct the direction of a low-complexity signal from the basis pursuit denoising program (BPDN), an \emph{instance optimal} algorithm whose reconstruction error is controlled by the restricted isometry property of the sensing matrix~\cite{candes2005decoding,Foucart_2013}.  Using this algorithm, we can bound the reconstruction error of the direction of a sparse signal observed in a PO-CS model. This error bound is \emph{(i)} \emph{non-uniform}, in the sense that the estimation is possible, with high probability (\whp), given the observed signal, and \emph{(ii)} \emph{stable and robust} as the instance optimality of BPDN allows for both a bounded noise on the observed measurement phases and an error on the modeling of the signal by a sparse vector. Numerically, we observe that the number of measurements required for robust estimation of signal direction from noiseless phase-only observation is about twice the one needed for signal estimation in the case of (linear)~CS. 
\medskip

\noindent{\bf Notations:} Hereafter, the values $C,D,c>0$ represent universal constants whose values vary from one instance to the other. 

\subsection{Preliminaries}
\label{sec:preliminaries}

In (complex) compressive sensing, a real signal $\bs x \in \bb R^n$ is sensed with a complex sensing matrix $\bs A = \bs A^\Re + \im \bs A^\Im \in \bb C^{m \times n}$ through the model~\cite{Foucart_2013}\vse
\begin{equation}
  \label{eq:cs-model}
  \bs y = \bs A \bs x + \bs \epsilon, \vse
\end{equation}
for some complex measurement noise $\bs \epsilon \in \bb C^m$. In this summary, we assume $\bs x$ in the set $\Sigma^n_s$ of $s$-sparse vectors (or is well approximated by one of its elements). Our approach extends to other low-complexity signal sets, with sample complexities related to their Gaussian width \cite{LJ20,baraniuk2010model,Ayaz_2016,fazel2002matrix}.

The signal $\bs x$ can be recovered from $\bs y$, if, for some $0<\delta<1$, $\bs A$ satisfies the restricted isometry property of order $2s$, or RIP$(\Sigma^n_{2s},\delta)$, defined by~\cite{candes2005decoding, Foucart_2013}\vse
\begin{equation}
\label{eq:RIP-def}
\ts   (1-\delta) \|\bs u\|^2 \leq \|\bs A \bs u\|^2 \leq (1+\delta) \|\bs u\|^2, \quad \forall \bs u \in \Sigma^n_{2s}. \vse
\end{equation}
For instance, this property holds \whp if $\bs A = \bs \Phi / \sqrt{2m}$, with $\bs \Phi \sim \cl N^{m \times n}_{\bb C}(0,2)$ a complex $m \times n$ random Gaussian matrix such that $\Phi_{kl} \sim_{\iid} \cl N(0,1) + \im \cl N(0,1)$, and if $m \geq C \delta^{-2} s \ln({n}/{\delta s})$~\cite{Foucart_2013,baraniuk2008simple}.   

If $\bs A$ respects the RIP$(\Sigma^n_{2s},\delta)$ with $\delta < 1/\sqrt 2$~\cite{cai2013sparse} (see also~\cite[Thm 6]{foucart2016flavors}), then, for $\varepsilon \geq \|\bs \epsilon\|$, the \emph{basis pursuit denoising}~\cite{chen2001atomic} estimate $\hat{\bs x} = \Delta(\bs y, \bs A; \varepsilon)$ with \vse
\begin{equation}
  \label{eq:BP-def}
\ts  \Delta(\bs y, \bs A; \varepsilon) := \argmin_{\bs u} \|\bs u\|_1 \st \|\bs A \bs u - \bs y\| \leq \varepsilon, \vse \end{equation}
satisfies the \emph{instance optimality} relation \vse
\begin{equation}
  \label{eq:l2-l1-inst-opt}
\ts \|\bs x - \hat{\bs x}\| \leq C s^{-1} \|\bs x - \bs x_s\|_1 + D \varepsilon, \vse
\end{equation}
with $\bs x_s$ the closest $s$-sparse signal to $\bs x$ (that is, its best $s$-sparse approximation). Consequently, if $\bs x \in \Sigma^n_s$, $\bs \epsilon=\bs 0$, and $\varepsilon=0$, we get perfect signal recovery ($\hat{\bs x} = \bs x$).

We focus this work on the use of BPDN but our approach extends to any other instance optimal algorithm verifying \eqref{eq:l2-l1-inst-opt} \cite{LJ20}, such as orthogonal matching pursuit (OMP), compressive sampling matching pursuit (CoSaMP), or iterative hard thresholding (IHT)~\cite{Tropp_2007,blumensath2009iterative,foucart2016flavors,needell2009cosamp}

\subsection{Phase-only Compressive Sensing}
\label{sec:second-section}

Departing from linear CS, we now consider the phase-only compressive sensing (PO-CS) model inspired by \cite{Oppenheim_1981,boufounos2013sparse}, \vse
\begin{equation}
  \label{eq:PO-CS}
  \bs z = \signc (\bs A \bs x) + \bs \epsilon, \vse
\end{equation}
with $\signc \lambda := \lambda /|\lambda|$ if $\lambda \in \bb C \setminus \{0\}$ and $\signc 0 := 0$, and $\bs \epsilon \in \bb C^{m}$ a bounded noise with $\|\bs \epsilon\|_\infty \leq \tau$ for some $\tau \geq 0$.

We first focus on the noiseless case, $\bs \epsilon = 0$. Since the signal amplitude is non-observable in the model~\eqref{eq:PO-CS} (as in one-bit CS \cite{plan2012robust}), we consider the recovery of the normalized vector  
\begin{equation}
  \label{eq:norm-hyp}%\label{eq:def-norm-x}
\ts  \bs \nx :=  \frac{\kappa \sqrt m}{\|\bs A \bs x\|_1} \bs x,\ {\rm with}\ \|\bs A  \bs \nx\|_1= \kappa \sqrt m,\ \kappa := \sqrt{\frac{\pi}{2}}. \vse
\end{equation}
%%%%
% Since \eqref{eq:PO-CS} is invariant under a rescaling of $\bs x$ (as in one-bit CS \cite{plan2012robust}), we set \vse
% \begin{equation}
%   \label{eq:norm-hyp}
%   \|\bs A  \bs x\|_1=\kappa \sqrt m,\ \text{with}\ \kappa := \sqrt{\pi/2}. \vse
% \end{equation}
Since for $\bs A = \bs \Phi/\sqrt m$ with $\bs \Phi \sim \cl N^{m \times n}_{\bb C}(0,2)$, $\bb E \|\bs A \bs x\|_1=\kappa \sqrt m \|\bs x\|$  \cite[Lem. 5.2]{LJ20}, $\|\bs \nx\| \approx 1$ for such a random matrix.

Following~\cite{boufounos2013sparse}, we recast the PO-CS model as a linear CS model. Introducing $\bs \alpha_{\bs z} := \bs A^* \bs z/(\kappa \sqrt m) \in \bb C^m$, we see that the \emph{phase-consistency} and normalization constraints defined by $\diag(\bs z)^* \bs A \bs u \in \bb R^m_+$ and $\scp{\bs \alpha_{\bs z}}{\bs u} = 1$, respectively, are respected for $\bs u = \bs \nx$. Since $\bs \nx$ is real, this is equivalent to \vse
\begin{equation}
  \label{eq:consistency-real-imag}
  \begin{cases}
        \bs H_{\bs z} \bs u = \bs 0,\quad \scp{\bs \alpha^\Re_{\bs z}}{\bs u} = 1,\quad \scp{\bs \alpha^\Im_{\bs z}}{\bs u} = 0,\\
    (\bs D_{\bs z}^\Re \bs A^\Re + \bs D_{\bs z}^\Im \bs A^\Im ) \bs u > \bs 0,
  \end{cases}\vse
\end{equation}
with $\bs u \in \bb R^n$, and $\bs D_{\bs v} := \diag(\bs v)$, $\bs H_{\bs v} := \Im(\bs D_{\bs v}^* \bs A) = \bs D_{\bs v}^\Re \bs A^\Im - \bs D_{\bs v}^\Im \bs A^\Re $ for $\bs v \in \bb C^m$.

A meaningful estimate $\hat{\bs x} \in \bb R^n$ of $\bs \nx$ should at least respect the constraints \eqref{eq:consistency-real-imag}. We can relax them by discarding the second line of \eqref{eq:consistency-real-imag} and rather impose our estimate to respect \vse
\begin{equation}
  \label{eq:equiv-cs-model}
\bs A_{\bs z} \bs u = \bs e_1 := (1, 0,\, \cdots, 0)^\top = \bs A_{\bs z} \bs \nx,  \vse
\end{equation}
with, for $\bs v \in \bb C^m$, $\bs A_{\bs v} := ( \bs \alpha^\Re_{\bs v}, \bs \alpha^\Im_{\bs v}, \bs H^\top_{\bs v})^\top \in \bb R^{(m+2)\times n}$.

We note that \eqref{eq:equiv-cs-model} holds for $\bs u = \bs \nx$, since $\|\bs A \bs \nx\|_1 = \kappa \sqrt m$ and $\bs H_{\bs z} \bs x = \bs 0$. This equation thus stands for the constraint of an equivalent linear CS recovery problem.

Therefore, following Sec.~\ref{sec:preliminaries}, provided that $\bs A_{\bs z}$ respects the RIP$(\Sigma^n_{2s}, \delta)$ with $0 < \delta < 1/\sqrt 2$ (as shown in the next theorem), we can recover the direction of $\bs x \in \Sigma^n_s$ from $\bs z = \signc(\bs A \bs x) = \signc(\bs A \bs \nx)$ and, \ie $\hat{\bs x} = \Delta(\bs e_1, \bs A_{\bs z}; 0) = \bs \nx$.
\begin{theorem}[{Adapted to $\Sigma^n_s$ from \cite[Thm 3.1]{LJ20}}]
  \label{thm:rip-for-Ax}
  Given $0<\delta <1$, $\bs A = \bs \Phi/\sqrt m$ with $\bs \Phi \sim \cl N_{\bb C}^{m \times n}(0,2)$, $\bs x' \in \bb R^n$, and $\bs A_{\bs z'}$ defined above from $\bs A$ and $\bs z' = \signc(\bs A \bs x')$, if $m \geq C \delta^{-2} s \ln(n/s)$, then, with probability exceeding $1 - C \exp(-c \delta^2 m)$, $\bs A_{\bs z'}$ satisfies the RIP$(\Sigma^n_s,\delta)$. 
\end{theorem}
\noindent The proof relies on combining the RIP of $\bs A$ (to condition $\bs H_z$), and a variant of the RIP --- relating $\ell_2$-norm of sparse vectors to the $\ell_1$-norm of their projections \cite{foucart2016flavors} --- to bound $|\scp{\bs \alpha_{\bs z}}{\bs u}|$ for any sparse vectors $\bs u$~\cite{LJ20}. 
\medskip

For noisy PO-CS, we must adapt \eqref{eq:equiv-cs-model} as $\bs z$ is corrupted by $\bs \epsilon$. Interestingly, as explained hereafter, $\bs u = \bs \nx$ respects   \vse
$$
\|\bs A_{\bs z} \bs u - \bs e_1\| \leq \varepsilon, \vse
$$
a noisy version of the previous fidelity constraint. In fact, for $\bs A = \bs \Phi/\sqrt m$, $\bs \Phi \sim \cl N^{m \times n}_{\bb C}(0,2)$, and $\|\bs \epsilon\|_\infty \leq \tau$, we can prove that, given $\delta > 0$, \vse
$$
  \ts \|\bs A_{\bs z} \bs \nx - \bs e_1\| \leq \varepsilon(\tau) := \sqrt 2 \tau \frac{1+\delta}{1-\delta} \vse
$$
with probability exceeding $1-C\exp(-c \delta^2 m)$. Moreover, provided that $0<\delta + 9 \tau < 1$ and $m \geq C (1+\delta^{-2}) s \ln(n/s)$, $\bs A_{\bs z}$ satisfies \whp the RIP$(\cl K, \delta + 9\tau)$ \cite[Thm 4.3]{LJ20}. In this case, the estimate $\hat{\bs x} = \Delta(\bs e_1, \bs A_{\bs z}, \varepsilon)$ respects \whp (\ref{eq:l2-l1-inst-opt}) (with $\bs x \to \bs \nx$). This proves that, provided $\bs \epsilon$ has bounded amplitude, with small variations compared to unity, one can stably and robustly estimate the direction of $\bs x$ (as encoded in $\bs \nx$). 

\subsection{Numerical Experiments}
\label{sec:experim}

\begin{figure}[t]  
    \centering
    \includegraphics[height=2.8cm]{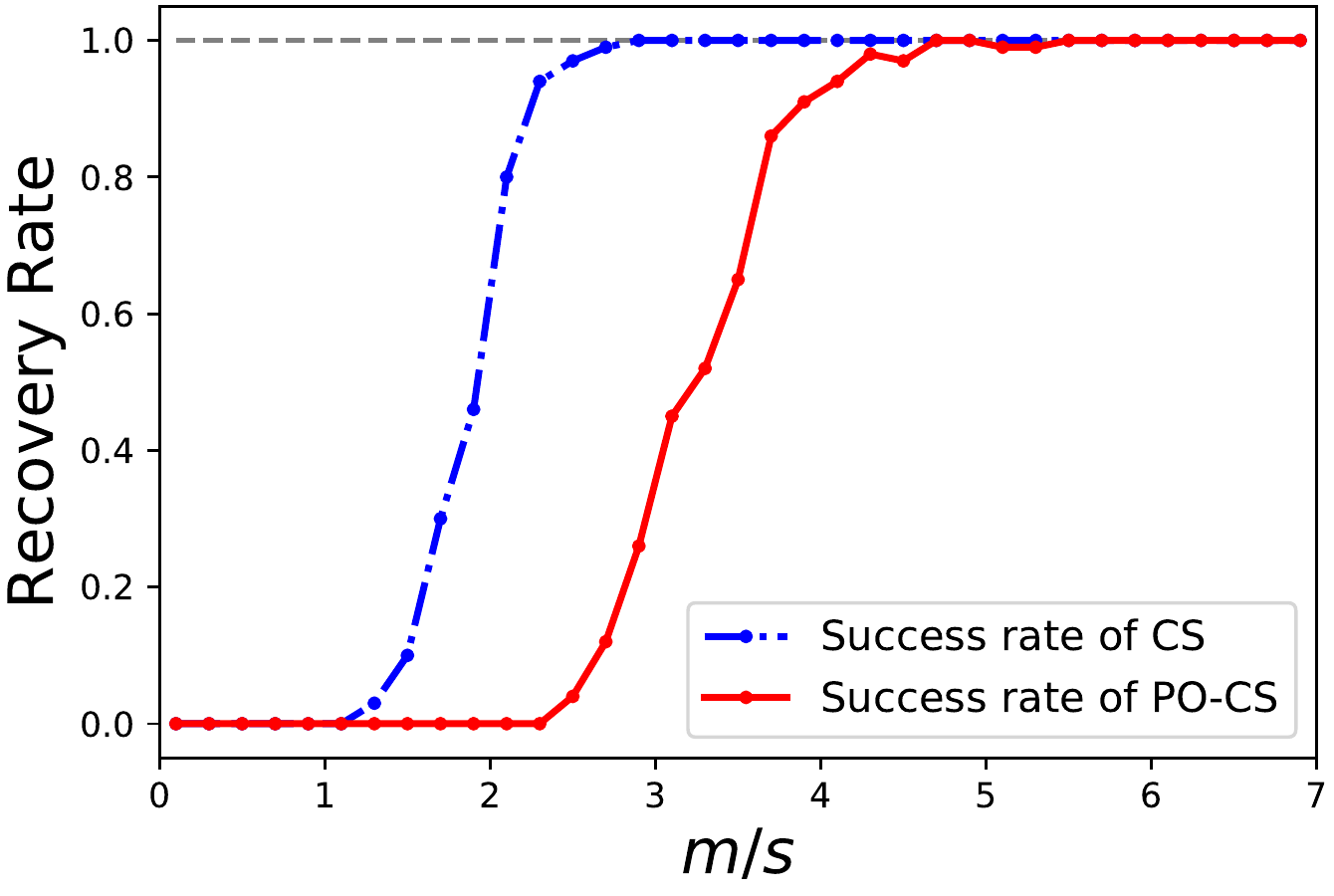}\hspace{0.01\columnwidth}
    \includegraphics[height=2.83cm]{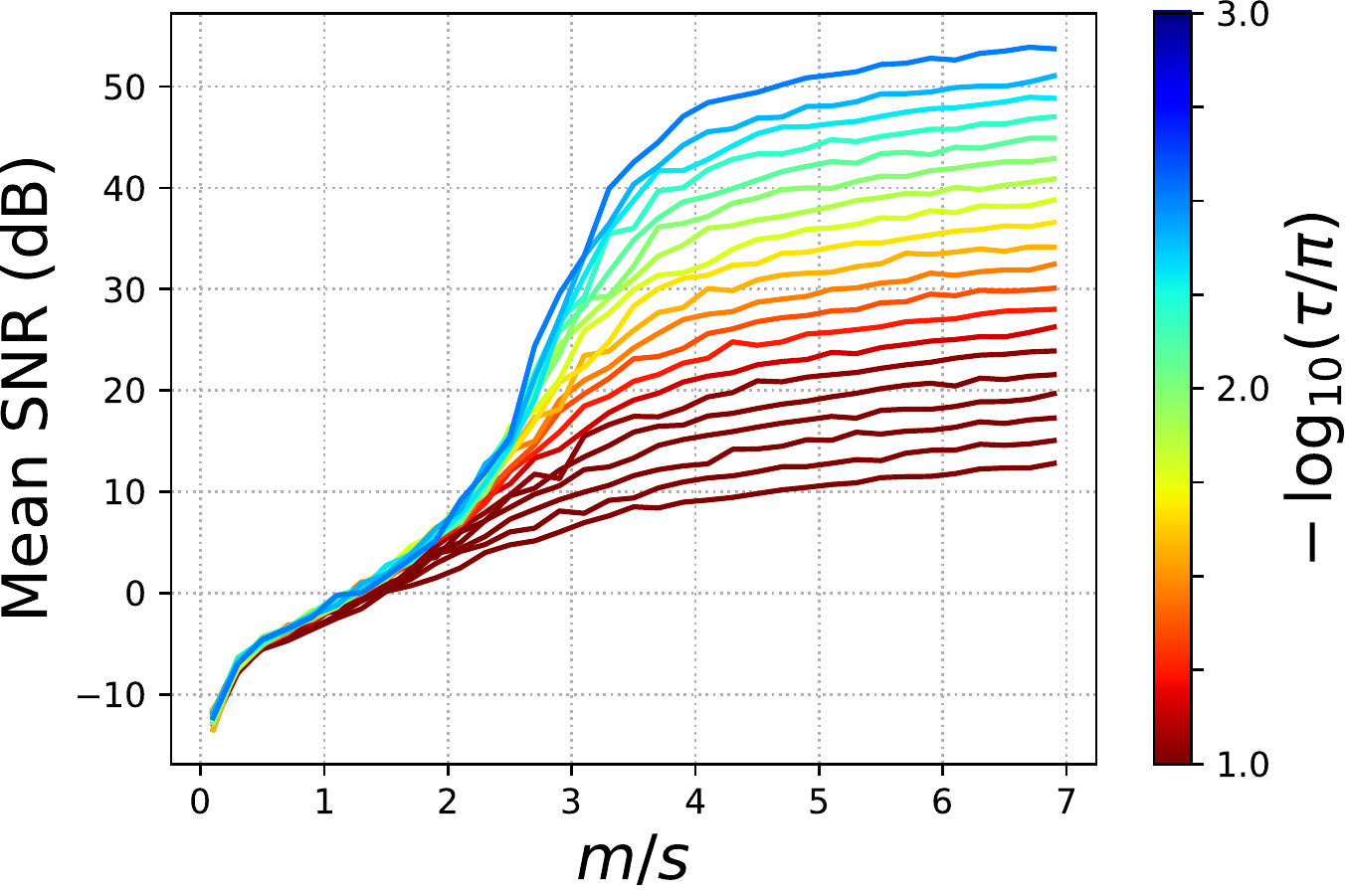} \vspace{-3mm}
    \caption{(left) Rate of successful recovery of $s$-sparse signals in function of $m/s$ for PO-CS (in red) under the hypothesis \eqref{eq:norm-hyp}, and in linear CS (dashed blue). (right, best viewed in color) Average SNR in function of $m/s$ for the estimation of the direction of $s$-sparse signals in PO-CS. The color coding of the SNR curves corresponds to the noise level $\tau$, as measured by $-\log_{10} \tau/\pi \in [1, 3]$. All figures were generated by Matplotlib~\cite{Hunter:2007}.\vspace{-4mm}}
    \label{fig:BP_rec_rate}
\end{figure}

Let us test the recovery of the direction of sparse vectors in PO-CS, both for noiseless and noisy sensing models. 

First, we compare in a noiseless scenario the sample complexities of both PO-CS and (linear) CS for sparse signal recovery. We have thus randomly generated $(s=10)$-sparse vectors in $\bb R^{(n=100)}$ (random support, random Gaussian entries) for a range of measurements $m \in [1, 70]$. For each $m$, to reach valid statistics, we generated 100 different $\bs \Phi \sim \cl N_{\bb C}^{m \times n}(0,2)$ and sparse signals $\bs x$, each normalized so that $\bs x = \bs \nx$ in (\ref{eq:norm-hyp}).

Numerically, both PO-CS (when computing $\hat{\bs x} = \Delta(\bs A_{\bs z}, \bs e_1; 0)$) and CS (for $\hat{\bs x} = \Delta(\bs A, \bs A \bs x; 0)$) were solved with BPDN, using the Python-SPGL1 toolbox\footnote{\url{https://github.com/drrelyea/spgl1}}~\cite{van_den_Berg_2009}, together with the Numpy module~\cite{van2011numpy}. A reconstruction was considered successful when $\|\bs x - \hat{\bs x}\|/\|\bs x\| \leq 10^{-3}$ (\ie a 60\,dB SNR). 

Fig.~\ref{fig:BP_rec_rate}(left) displays the success rate of both approaches, as measured from the fraction of successful reconstructions over the 100 trials in function of $m$. PO-CS requires about twice the number of measurements needed for perfect signal recovery in linear CS. This is naively expected as the model \eqref{eq:PO-CS} provides $m$ constraints (associated with $m$ phases) compared to \eqref{eq:cs-model} that delivers $2m$ independent observations.  

Second, the noisy model (\ref{eq:PO-CS}) is considered with the same experimental setup as above, with $\epsilon_k \sim_\iid \cl U( \tau \cl B)$ for $k \in [m]$, $\cl B = \{\lambda \in \bb C: |\lambda| \leq 1\}$. The noise level is compared to $\pi$ as reconstructing the signal direction is (in general) impossible for $\tau \geq \pi$ \cite{LJ20}. 
Fig.~\ref{fig:BP_rec_rate}(right) shows the average signal-to-noise ratio (SNR) $20 \log_{10} \|\bs x\|/\|\bs x - \hat{\bs x}\|$ as a function of $m/s$, colored by the value of $\tau$. From $m/s \simeq 3$, the SNR level grows linearly with the increase of $-\log_{10} \tau/\pi$, as involved by taking the logarithm of both sides of \eqref{eq:l2-l1-inst-opt}.

\subsection{Conclusion}
\label{sec:conclusion}

This work proved that the direction of a sparse signal can be perfectly reconstructed from the phase of its complex Gaussian random projections. This is achieved from an equivalent linear sensing model combining a signal normalization and a phase-consistency constraints.  The basis pursuit denoising program can then be used for a practical reconstruction. The produced estimate is also both robust to measurement noise (provided the noise amplitude is bounded) and stable with respect to modeling error, if the signal is not exactly sparse. 

When implemented with the CS paradigm, applications such as radar, magnetic resonance imaging, or computed tomography, rely on the random subsampling of the signal frequency domain. This is associated with a structured complex random sensing matrix. In these contexts, PO-CS is appealing since PO measurements are insensitive to large amplitude variations. Following~\cite{Oppenheim_1982,Oppenheim_1981}, a critical open problem is to extend the approach to partial random Fourier matrices or to other structured random sensing constructions. A first breakthrough in this direction would be to verify this extension for partial Gaussian circulant matrices, which are applicable to one-bit CS~\cite{Dirksen_2019}.

\subsection*{Acknowledgments}

Part of this research was supported by the Fonds de la Recherche Scientifique -- FNRS under Grant T.0136.20 (Project Learn2Sense). 

%% You can make the bibliography smaller

\end{document}